\documentclass[lettersize, journal]{IEEEtran}
\usepackage{amsmath, amsfonts}
\usepackage{algorithmic}
\usepackage{algorithm}
\usepackage{array}
\usepackage{subfigure}
\usepackage{caption}
\usepackage{textcomp}
\usepackage{stfloats}
\usepackage{url}
\usepackage{verbatim}
\usepackage{graphicx}
\usepackage{cite}
\usepackage{color}
\usepackage{bbm}
\usepackage{amssymb, amsmath, amsthm, amsfonts}
\newtheorem{lemma}{Lemma}
\allowdisplaybreaks
\begin{document}

\title{Performance Analysis of Pinching-Antenna-Enabled Internet of Things Systems}
%(DEIN)Integrated Data and Energy Communication Networks
\author{Han~Zhang,~\IEEEmembership{Student Member,~IEEE,}
	Bingxin~Zhang,~\IEEEmembership{Member,~IEEE,}
	Yizhe~Zhao,~\IEEEmembership{Member,~IEEE,}
	Kun~Yang,~\IEEEmembership{Fellow,~IEEE,}
	and~Guopeng~Zhang% <-this % stops a space
	%\thanks{This paper was partly funded by Jiangsu Major Project on Fundamental Researches (Grant No.: BK20243059), Gusu Innovation Project for People (Grant No.: ZXL2024360), Natural Science Foundation of China (Grant No. 62132004) and Nanjing University-China Mobile Communications Group Co.,Ltd. Joint Institute. (\textit{Corresponding author: Kun Yang.})}
	% \thanks{Bingxin Zhang is with the School of Information and Communication Engineering, University of Electronic Science and Technology of China, Chengdu 611731, China (e-mail: bxzhang@std.uestc.edu.cn).}
	\thanks{Han Zhang, Bingxin Zhang and Kun Yang  are with the State Key Laboratory of Novel Software Technology, Nanjing University, Nanjing 210008, China, and School of Intelligent Software and Engineering, Nanjing University (Suzhou Campus), Suzhou, 215163, China (email: hanzhangl@smail.nju.edu.cn; bxzhang@nju.edu.cn; kunyang@nju.edu.cn).}
	\thanks{Yizhe Zhao is  with the School of Information and Communication Engineering, University of Electronic Science and Technology of China, Chengdu 611731, China (e-mail: yzzhao@uestc.edu.cn).}
	\thanks{Guopeng Zhang is with the School of Computer Science and Technology, China University of Mining and Technology, Xuzhou 221116, China (e-mail: gpzhang@cumt.edu.cn).}
}

%\markboth{Journal of \LaTeX\ Class Files, ~Vol.~14, No.~8, August~2021}%
%\IEEEpubid{0000--0000/00\$00.00~\copyright~2023 IEEE}

\maketitle

\begin{abstract}
The pinching-antenna systems (PASS), which activate small dielectric particles along a dielectric waveguide, has recently emerged as a promising paradigm for flexible antenna deployment in next-generation wireless communication networks. While most existing studies assume rectangular indoor layouts with full coverage waveguide, practical deployments may involve geometric constraints, partial coverage, and non-negligible waveguide attenuation. This paper presents the first analytical investigation of PASS in a circular indoor environment, encompassing both full coverage and partial coverage waveguide configurations with/without propagation loss. A unified geometric–propagation framework is developed that jointly captures pinching-antenna placement, Internet of Things (IoT) device location distribution, and waveguide attenuation. Closed-form expressions for the outage probability and average achievable rate are derived for four scenarios, with accuracy validated via extensive Monte-Carlo simulations. The analysis reveals that, under the partial coverage waveguide scenario with propagation loss, the system performance demonstrates a non-monotonic trend with respect to the waveguide length, and the optimal length decreases as the attenuation coefficient increases. Numerical results further quantify the interplay between deployment strategy, waveguide propagation loss, and coverage geometry, offering practical guidelines for performance-oriented PASS design.
\end{abstract}
\begin{IEEEkeywords}
Pinching-Antenna Systems (PASS), Dielectric Waveguides, Outage Probability, Average Achievable Rate
\end{IEEEkeywords}

\section{Introduction}
% \IEEEPARstart{I}{n} the future communication networks, comprehensive monitoring and sensing of the real world will be achieved through various sensors 

{\huge T}he rapid proliferation of the Internet of Things (IoT) is driving an unprecedented demand for flexible, scalable, and energy-efficient wireless technologies to support massive device connectivity and diverse application requirements. Emerging IoT services, such as smart healthcare, industrial automation, intelligent transportation, and immersive extended reality, call for communication systems that can dynamically adapt to heterogeneous deployment scenarios, provide reliable links for densely distributed devices, and operate efficiently under stringent energy and latency constraints \cite{tataria20216g, wang2023road}. As IoT networks continue to evolve, critical challenges such as spectrum scarcity, interference management, and severe signal degradation in high-frequency bands are becoming increasingly pronounced \cite{saad2019vision}. Flexible antenna systems capable of reconfiguring the wireless environment have therefore emerged as promising enablers for future IoT connectivity. By dynamically adjusting antenna positions, orientations, or geometrical configurations \cite{yang2025flexible}, such systems can enhance coverage, improve reliability, and optimize resource utilization in highly dynamic IoT environments.

Several flexible antenna technologies have been proposed in recent years. Reconfigurable intelligent surfaces (RIS) \cite{pan2021reconfigurable, huang2019reconfigurable, di2020smart, chen2022reconfigurable} and intelligent reflecting surfaces (IRS) \cite{zhang2025performance, wu2019intelligent} have attracted significant attention for their ability to manipulate the wireless propagation environment by intelligently reflecting and refracting signals. By controlling the reflection coefficients of a large number of passive elements, RIS/IRS can enhance spatial diversity, extend coverage, improve energy efficiency, and reduce interference \cite{wu2019towards}. The development of simultaneously transmitting and reflecting RIS (STAR-RIS) \cite{mu2021simultaneously} further broadens RIS/IRS applicability by enabling simultaneous bidirectional signal control, making it suitable for more complex scenarios. However, RIS/IRS-assisted communication suffers from the cascaded effect of two large-scale fading components, one from the transmitter to the RIS/IRS and another from the RIS/IRS to the receiver, which can significantly attenuate the end-to-end signal power in long-range deployments.

Distinct from RIS/IRS that reconfigures the wireless environment between the transmitter and receiver, fluid antennas and movable antennas are directly embedded in the transceiver to flexibly adapt the signal transmission and reception process. Fluid antennas utilize liquid-based conductive materials to dynamically alter their shape, size, and position \cite{wong2020performance, new2024tutorial,wong2020fluid}. This allows fluid antennas to reconfigure their radiation characteristics, providing enhanced adaptability for scenarios with dynamic fading and interference patterns. The concept of fluid antenna multiple access (FAMA) \cite{wong2021fluid} exploits this flexibility to support multiple users with improved spectral efficiency. Movable antennas \cite{zhu2023movable, zhu2023modeling, chen2024movable} further enhance adaptability by physically repositioning the entire antenna structure in real time to optimize the link quality. While fluid antennas and movable antennas offer notable advantages in combating small-scale fading and improving local coverage, their ability to mitigate large-scale path loss remains limited, especially in extended-distance or blockage-prone environments.

%A key research problem in PASS is determining the optimal PA positions to minimize large-scale path loss and ensure constructive signal combination. Recent studies have explored multiple aspects of PASS.
To overcome these limitations, researchers have recently proposed novel flexible antenna architectures that can effectively mitigate the performance degradation caused by large-scale fading. Pinching-antenna systems (PASS) have thus emerged as a compelling alternative. PASS deploys pinching-antenna (PA) along a dielectric waveguide, enabling dynamic activation at different locations to optimize the wireless link adaptively. In \cite{xu2025joint}, a majorization–minimization and penalty dual decomposition algorithm was developed to address nonconvex beamforming optimization by decoupling system parameters.
The authors in \cite{tegos2025minimum} examined uplink PASS minimum-rate maximization under different user distributions, where an iterative algorithm was designed to jointly optimize antenna positions and beamforming for user fairness.
Antenna activation in non-orthogonal multiple access (NOMA) assisted PASS was investigated in \cite{wang2025antenna}, where a low-complexity optimization algorithm was shown to significantly improve sum rate and spectral efficiency.
Deep learning-based channel estimation approaches were presented in \cite{xiao2025channel} to enhance estimation accuracy while reducing pilot overhead.
Both multi-waveguide and single-waveguide deployments were analyzed in \cite{ding2025flexible}, demonstrating that integrating PASS with NOMA can further boost data rates and spectral efficiency.
Closed-form outage probability and average rate expressions, incorporating waveguide propagation loss and antenna placement, were derived in \cite{tyrovolas2025performance}.
The uplink performance of PASS with multiple pinching-antennas serving a single user was analyzed in \cite{hou2025performance}, highlighting considerable throughput gains.
In \cite{cheng2025performance}, the impact of PA position optimization under orthogonal multiple access(OMA) and NOMA schemes was studied, and effective throughput enhancement strategies were proposed.
Moreover, \cite{ouyang2025array} provided an analytical characterization of PASS array gain, revealing the influence of antenna number and spacing on system performance.

These works collectively highlight the ability of PASS in various scenarios, yet despite these advancements, most existing PASS studies consider rectangular indoor environments where the dielectric waveguide is assumed to be sufficiently long to cover the entire target region. Under this full coverage assumption, the performance implications of limited waveguide coverage and realistic waveguide attenuation remain underexplored. Moreover, the effect of geometric constraints, such as circular building layouts (e.g., arenas, stadiums, round theaters, and iconic landmarks), or partial coverage waveguide deployments, has not been analytically characterized in the literature.

Motivated by the aforementioned research gaps, this paper makes the first attempt to investigate PASS in a circular indoor environment by developing two distinct geometric propagation models: a full coverage waveguide and a partial coverage waveguide. The proposed unified framework jointly incorporates PA deployment strategies and physical waveguide attenuation into a single channel model, and derives closed-form expressions for four representative configurations full/partial coverage with/without waveguide propagation loss, thereby providing useful insights. 
%Notably, for the partial coverage waveguide with propagation loss case, the analysis reveals a non-monotonic relationship between waveguide length and system performance: short lengths suffer from coverage deficiency, while long lengths incur excessive attenuation, leading to performance degradation. The optimal waveguide length is shown to be affected by the attenuation coefficient, providing practical guidelines for length optimization in PASS.

The main contributions of this paper are summarized as follows:
\begin{enumerate}
\item A novel system model is developed for PA-enabled downlink wireless communication in a circular indoor environment. Unlike previous works restricted to rectangular layouts with full-coverage waveguides, our model accounts for both full and partial coverage, together with dynamic PA placement determined by the user device location.
\item A comprehensive analytical framework is established to derive closed-form expressions for outage probability and average achievable rate under four configurations full/partial coverage waveguide with/without propagation loss jointly accounting for geometric propagation, deployment strategy, and waveguide attenuation.
\item In the partial coverage waveguide with propagation loss case, the analysis uncovers a non-monotonic dependence of performance on waveguide length and shows that the optimal length is affected by the attenuation coefficient, offering actionable guidelines for practical deployments. The accuracy of the analytical results is further validated through extensive simulations.
\end{enumerate}

The remainder of this paper is organized as follows. Section II introduces the system model. Section III presents the performance analysis for full coverage waveguide with/without propagation loss. Section IV extends the analysis to partial coverage configurations, where closed-form expressions for the outage probability and the average achievable rate are derived for both cases. Section V provides numerical results to validate the analytical expressions and extract design insights. Finally, Section VI concludes the paper.

\section{System model} \label{sec:SM}

\begin{figure}[!t]
\centering
\includegraphics[width=3.3 in]{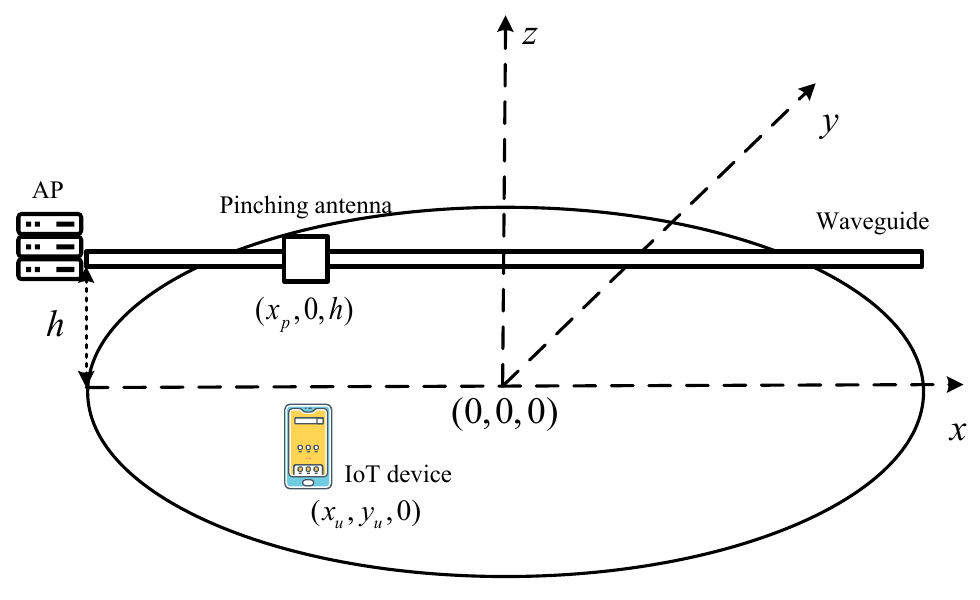}
\caption{The pinching-antenna system.}
\label{sysmod}
\end{figure}

We investigate a PA-enabled downlink wireless communication system deployed inside a circular building, as depicted in Fig.~\ref{sysmod}. The system comprises an access point (AP), a pinching-antenna mounted on a waveguide, and a single-antenna internet of things (IoT) device located within the room. Let $r$ denote the radius of the circular region. The AP is positioned at $(-r, 0, h)$, i.e., at a height $h$ above the boundary of the circular area, and is connected to the waveguide. The waveguide, with length equal to the diameter $2r$ of the circular area, is horizontally installed such that its projection passes through the center of the building. Thus, the ends of the waveguide are located at $(-r, 0, h)$ and $(r, 0, h)$. The PA is attached to the waveguide at a configurable position $(x_p, 0, h)$, enabling dynamic adjustment of the transmission path according to the IoT device's location.

In this setup, the IoT device situated at $\psi_u = (x_u, y_u , 0)$ is assumed to be uniformly distributed within the circular region, i.e., the joint probability density function (PDF) of $(x_u, y_u)$ is
\begin{align}
f_{x_u, y_u}(x, y) = 
\begin{cases}
    \frac{1}{\pi r^2}, & x^2 + y^2 \leq r^2, \\[8pt]
    0, & \text{otherwise}.
\end{cases}
\end{align}

The PA dynamically enhances the wireless link by adjusting its position along the waveguide. This architecture ensures efficient coverage within the circular region by optimizing the transmission path based on the IoT device's location. Let $\psi_p = (x_p, 0, h)$ denote the position of the PA with $x_p \in [-r, r]$, the channel between the PA and the IoT device can be expressed as \cite{tyrovolas2025performance}
\begin{align}
    h_{p, u} = \frac{\sqrt{\eta} e^{-j \frac{2\pi}{\lambda} |\psi_p - \psi_u|}}{|\psi_p - \psi_u|},
\end{align}
where $\eta = \frac{c^2}{16 \pi^2 f_c^2}$ denotes the free-space path loss factor at a reference distance of 1~m, with $c$ being the speed of light and $f_c$ representing the carrier frequency. $|\psi_p - \psi_u|$ denotes the distance between the PA and the IoT device, and can be expressed as
\begin{align}
    |\psi_p - \psi_u| = \sqrt{h^2 + y_u^2 + (x_u - x_p)^2}.
\end{align}

However, as the signal propagates through the waveguide, its phase evolves according to the propagation properties of the dielectric medium. In particular, the interaction between the signal and both the dielectric core and its surrounding material decreases the phase velocity, which is quantified by the effective refractive index $n_{eff}$. This reduction in phase velocity leads to a guided wavelength of $\lambda_g = \frac{\lambda}{n_{eff}}$. As a result, the signal radiated by the PA accumulates an additional phase shift, which can be expressed as \cite{tyrovolas2025performance}
\begin{align}
    h_{a, p} = e^{-j \frac{2 \pi}{\lambda_g}|\psi_p - \psi_a|},
\end{align}
where $\boldsymbol{\psi}_a = (0, 0, h)$ specifies the position of the waveguide feed point. $|\psi_p - \psi_a|$ denotes the propagation distance of the signal within the waveguide, and can be expressed as
\begin{align}
    |\psi_p - \psi_a| = x_p + r.
\end{align}

The waveguide exhibits an absorption coefficient $\alpha \in [0, +\infty)$, which models the intrinsic power loss experienced by the signal during propagation. Accordingly, under the assumption of exponential attenuation, the signal received by the IoT device can be written as
\begin{align}
    y_{r}=\sqrt{P_{t} e^{-\alpha\left\|\psi_{p}-\psi_{a}\right\|}} h_{p, u} h_{a, p} s+\omega_{n},
\end{align}
where $P_t$ denotes the AP transmit power, and $s$ is the transmitted signal with unit power, i.e., $\mathbb{E}[|s|^2]=1$, where $\mathbb{E}[\cdot]$ denotes statistical expectation. $\omega_n$ represents additive white Gaussian noise (AWGN) with zero mean and variance $\sigma^2$. Accordingly, the received  signal-to-noise ratio (SNR) of the IoT device can be expressed as
\begin{align}
    \gamma &=\frac{\eta P_{t} e^{-\alpha | \psi_{p}-\psi_{a}|}\left| e^{-j (\frac{2\pi}{\lambda} |\psi_p - \psi_u| + \frac{2\pi}{\lambda_g} |\psi_p - \psi_a|)} \right|}{\sigma^{2}|\psi_{u}-\psi_{p}|^{2}} \notag\\
    & = \frac{\eta P_{t} e^{-\alpha (x_p + r)}}{\sigma^{2}\left(h^2 + y_u^2 + (x_u - x_p)^2\right)}.
\end{align}
\section{Performance analysis of downlink transmission}
We present the analytical derivations of key performance metrics, including the outage probability and the average achievable rate, for the PASS. The analysis assumes that the PA is positioned to minimize its distance to the IoT device, thereby enabling a comprehensive evaluation of the effectiveness of this placement strategy. Specifically, for the PA deployment strategy, we derive the key performance metrics for both the ideal case of a lossless dielectric waveguide and the practical case where the propagation loss of the waveguide is taken into account.
\subsection{full coverage waveguide without propagation loss}
We consider the case where the PA is dynamically located at $x_p = x_u$, corresponding to the minimum possible distance to the IoT device. This configuration effectively optimizes the large-scale fading between the pinching-antenna and the IoT device. When the waveguide propagation loss is neglected, the received SNR provides an upper bound on the PA performance, as it eliminates the attenuation introduced by waveguide propagation. The received SNR can then be rewritten as
\begin{align}
    \gamma = \frac{\eta P_{t}}{\sigma^{2}(y_u^2+h^2)}.
\end{align}

The outage probability can be defined as
\begin{align}
    P_{\text {out }} &=\operatorname{Pr}\left(\frac{\eta P_{t}}{\sigma^{2}\left(y_{u}^{2}+h^{2}\right)} \leq \gamma_{\text {th}}\right) \notag \\
    &= \operatorname{Pr}\left(y_{u}^{2} \geq \frac{\eta P_{t}}{\sigma^{2}\gamma_{\text {th}}} -h^{2}\right),
\end{align}
where $\gamma_{\text {th}}$ is the SNR threshold. We can observe that the outage probability is decided by the distribution of $y_u$ which can be expressed as
\begin{align}
    F_{{y_{u}}}(x)=\left\{\begin{array}{ll}1, & x > r, \\ \frac{2\left(x \sqrt{r^{2}-x^2}+r^{2} \arcsin \left(\frac{x}{r}\right)\right)}  {\pi r^{2}}, & 0 \leq x \leq r, \\ 0, & x< 0.\end{array}\right.
\end{align}

Let $A =\frac{\eta P_{t}}{\sigma^{2}\gamma_{\text {th}}} -h^{2}$. Then, the outage probability can be expressed as
\begin{align}
P_{\text {out }}=\left\{\begin{array}{ll}0, & A \geq r^2, \\ 1 - \frac{2\left(\sqrt{A} \sqrt{r^{2}-A}+r^{2} \arcsin \left(\frac{\sqrt{A}}{r}\right)\right)}  {\pi r^{2}}, & 0 \leq A \leq r^2, \\ 1, & A\leq 0.\end{array}\right.
\end{align}

When the waveguide propagation loss is negligible, the average achievable rate of the IoT device can be defined as
\begin{align}
    R_{p}=\mathbb{E}\left[\log _{2}\left(1+\frac{\eta P_{t}}{\sigma^{2}(y_u^2+h^2)}\right)\right].
\end{align}
\begin{lemma}
     The closed-form expression for the average achievable rate is given by
    \begin{align}
    R_{p} 
    &= \frac{1}{\ln 2} \ln\left(1 + \frac{\eta P_t}{\sigma^2 h^2}\right) + \frac{2}{\ln 2} \Bigg[
    \ln\left(\frac{1 + \Gamma}{1 + \Lambda}\right) 
    + \frac{1}{2}  \frac{1 - \Gamma}{1 + \Gamma} \notag \\
    &\qquad \qquad \qquad \qquad \qquad \qquad \qquad - \frac{1}{2}  \frac{1 - \Lambda}{1 + \Lambda}
    \Bigg],
	\end{align}
    where  $B \triangleq \eta P_t + \sigma^2 h^2, \Gamma \triangleq \sqrt{1 + \frac{\sigma^2 r^2}{B}}$ and $\Lambda \triangleq \sqrt{1 + \frac{r^2}{h^2}}$.
\end{lemma}
\begin{proof}
    Please refer to Appendix A.
\end{proof}

\subsection{full coverage waveguide with propagation loss}
When the propagation loss in the waveguide is considered, the received SNR can be expressed as
\begin{align}
    \gamma=\frac{\eta P_{t}e^{-\alpha(x_u+r)}}{\sigma^{2}(y_u^2+h^2)},
\end{align}
where $e^{-\alpha(x_p + r)}$ captures the additional power attenuation due to waveguide propagation loss. As a result, the outage probability can be expressed as
\begin{align}
    P_{\text {out }} &= \operatorname{Pr}\left(\gamma \leq \gamma_{\text {thr }} \right)= \operatorname{Pr}\left(y_{u}^{2} \geq \frac{\eta P_{t}e^{-\alpha(x_u+r)}}{\sigma^{2}\gamma_{\text {thr }}} -h^{2}\right) \notag \\
    & = \operatorname{Pr}\left(r^2 - x_{u}^{2} \geq \frac{\eta P_{t}e^{-\alpha(x_u+r)}}{\sigma^{2}\gamma_{\text {thr }}} -h^{2}\right).
\end{align}
Let $C =\frac{\eta P_{t}}{\sigma^{2}\gamma_{\text {thr }}}$, the integration region for the outage probability must satisfy $-\sqrt{r^2 - x_u^2} \leq y_u \leq -\sqrt{Ce^{-\alpha(x_u+r)}}$ or $\sqrt{Be^{-\alpha(x_u+r)}} \leq y_u \leq \sqrt{r^2 - x_u^2}$. To determine the integration region, it is necessary to analyze the solutions of the following equation
\begin{align}
    Ce^{-\alpha(x_u+r)} - h^2 = r^2 - x_u^2.
    \label{solution_num}
\end{align}

If (\ref{solution_num}) yields two distinct real solutions, denote them by $x_u = a$ and $x_u = c$, where $a < c$. It should be noted that, due to the monotonicity of the left-hand side expression $C e^{-\alpha(x_u + r)} - h^2$, the number of real solutions can be readily determined by straightforward calculation.If  (\ref{solution_num}) does not admit any real solution for $x_u$, the outage probability is $P_{out} = 0$.  In the case where there is a unique real solution $a$ within $[-r, r]$, let $x_u = b$ denote the root of  $C e^{-\alpha(x_u + r)} - h^2 = 0$. 

Then, we have
\begin{align}
    b = -\frac{\ln h^2 - \ln C}{\alpha} - r.
\end{align}

If $b \in [-r, r]$, we define $\rho(x_u) \triangleq \sqrt{r^2 - x_u^2}$ and $\Delta(x_u) \triangleq \sqrt{C e^{-\alpha(x_u + r)} - h^2}$. Then the outage probability can be expressed as
\begin{align}
P_{\mathrm{out}} &=
2 \int_{a}^{b} \int_{\Delta(x_u)}^{\rho(x_u)}
    \frac{1}{\pi r^2} dy_u dx_u 
+ 2 \int_{b}^{r} \int_{0}^{\rho(x_u)}
    \frac{1}{\pi r^2} dy_u dx_u \notag \\[2mm]
&= \frac{1}{\pi r^2}
    \left[
        r^2 \arcsin\left( \frac{x_u}{r} \right)
        + x_u \sqrt{r^2 - x_u^2}
    \right]_{a}^{r} \notag \\
    &\qquad \qquad \qquad \qquad \qquad \quad - \frac{2}{\pi r^2} \int_{a}^{b} \Delta(x_u) dx_u.
    \label{P_out_of_lossy_dielectric_waveguide}
\end{align}

To facilitate the integration of the second term in (\ref{P_out_of_lossy_dielectric_waveguide}), we first express \( x_u \) in terms of \( \Delta(x_u) \) as
\begin{align}
 x_u = -\frac{1}{\alpha} \ln\left( \frac{{\Delta(x_u)}^2 + h^2}{C} \right) - r.
\end{align}

The derivative of \( \Delta(x_u) \) with respect to \( x_u \) is given by
\begin{align}
    \frac{d\Delta(x_u)}{dx} = \frac{1}{2}  \frac{-\alpha C e^{-\alpha(x_u + r)}}{\sqrt{C e^{-\alpha(x_u + r)} - h^2}} 
    = \frac{-\alpha ({\Delta(x_u)}^2 + h^2)}{2\Delta(x_u)}.
\end{align}

Therefore, the differential \( dx_u \) can be expressed in terms of \( d\Delta(x_u) \) as
\begin{align}
dx = \frac{-2\Delta(x_u)}{\alpha({\Delta(x_u)}^2 + h^2)} \, d\Delta(x_u).
\label{Delta(x_u)}
\end{align}
Based on (\ref{Delta(x_u)}), the integral term in the second component of (\ref{P_out_of_lossy_dielectric_waveguide}) can be derived as
\begin{align}
\int_a^b \Delta(x_u)  dx &= \int_{\Delta(a)}^{\Delta(b)} \Delta(x_u)  \left( \frac{-2\Delta(x_u)}{\alpha({\Delta(x_u)}^2 + h^2)} \right) d\Delta(x_u) \notag\\
&= -\frac{2}{\alpha} \int_{\Delta(a)}^{\Delta(b)} \frac{{\Delta(x_u)}^2}{{\Delta(x_u)}^2 + h^2} \, d\Delta(x_u) \notag \\[2mm]
&= -\frac{2}{\alpha} \left[ \int_{\Delta(a)}^{\Delta(b)} 1  -  \frac{h^2}{{\Delta(x_u)}^2 + h^2} \, d\Delta(x_u) \right]\notag \\
&= -\frac{2}{\alpha} \left[ \Delta(x_u) - h \arctan\left( \frac{\Delta(x_u)}{h} \right) \right]_a^b. \notag \\[2mm]
\end{align}
Then let $\Phi(x) \triangleq h  \arctan\left( \frac{\Delta(x)}{h} \right) - \Delta(x)$, we have
\begin{align}
P_{\mathrm{out}} &=
\frac{1}{\pi r^2} \left[
    r^2 \left( \frac{\pi}{2} - \arcsin\left( \frac{a}{r} \right) \right)
    - a \sqrt{r^2 - a^2}
\right] \notag \\
&\quad - \frac{4}{\alpha \pi r^2} \left[ \Phi(b) - \Phi(a) \right].
\end{align}

If $a \in [-r, r]$ and $c \in [-r, r]$, we have
\begin{align}
P_{\mathrm{out}} =\ &
2\int_{a}^{c} \int_{\Delta(x_u)}^{\rho(x_u)} \frac{1}{\pi r^2} dy_u dx_u \notag \\[2mm]
=\ & 2 \int_{a}^{c} \frac{1}{\pi r^2} \left( \rho(x_u) - \Delta(x_u) \right) dx_u \notag \\[2mm]
=\ & \frac{2}{\pi r^2} \left[
\frac{1}{2} x_u \rho(x_u)
+ \frac{1}{2} r^2 \arcsin\left( \frac{x_u}{r} \right)
- \frac{2}{\alpha} \Phi(x_u)
\right]_a^c \notag \\[2mm]
=\ & \frac{1}{\pi r^2} 
r^2 \left( \arcsin\left( \frac{c}{r} \right) - \arcsin\left( \frac{a}{r} \right) \right)
+ \frac{1}{\pi r^2} c \rho(c)\notag \\
&- \frac{1}{\pi r^2} a \rho(a)
- \frac{4}{\alpha \pi r^2} \left[ \Phi(c) - \Phi(a) \right].
\end{align}

If $Be^{-\alpha(x_u+r)} - h^2 \leq r^2 - x_u^2$ for $x_u = -r$ and $x_u = r$, the outage probability is $P_{out} = 1$.

\begin{lemma}
When the waveguide propagation loss is taken into account, the average achievable rate can be expressed as
\begin{align}
R_{p} 
&\approx \pi r^2 \ln(\sigma^2 h^2) + 2\pi r^2 \left\{ \ln\left(\frac{1 + \Lambda}{2}\right)
+ \frac{1}{2}  \frac{1 - \Lambda}{1 + \Lambda} \right\}\notag \\
&\quad+ \frac{r\pi}{V} \sum_{m=1}^{V} \sqrt{1 - x_m^2} \Bigg[
    \frac{1}{\alpha} \mathrm{Li}_2\left( -\frac{\eta P_t}{\sigma^2(y_m^2 + h^2)} v_1 \right) \notag \\
    &\quad- \frac{1}{\alpha} \mathrm{Li}_2\left( -\frac{\eta P_t}{\sigma^2(y_m^2 + h^2)} v_2 \right)
\Bigg].
\end{align}
where $x_m = \cos\left( \frac{2m-1}{2V} \pi \right), y_m = r x_m, v = e^{-\alpha(x_u + r)}$, $v_1 = e^{\alpha \sqrt{r^2 - y_u^2} - \alpha r}$, $v_2 = e^{-\alpha \sqrt{r^2 - y_u^2} - \alpha r}$ and $\operatorname{Li}_2(\cdot)$ is the dilogarithm function.
\end{lemma}

\begin{proof}
    Please refer to Appendix B.
\end{proof}
\section{Extension to Partial Coverage Waveguide Scenarios}
In this section, we extend the system model to scenarios where the waveguide cannot fully cover the entire diameter. Such configurations correspond to practical environments, for example, spherical or dome-shaped structures such as planetariums, observatories, sports arenas, or large exhibition halls. The objective is to analyze the impact of insufficient waveguide coverage on communication performance. Specifically, this section first introduces the extended system model, and then derive closed-form expressions for the outage probability and the average achievable rate of the PASS under both scenarios with and without waveguide propagation loss.

% \begin{figure}[!t]
% \centering
% \includegraphics[width=3.3 in]{Problem_scenario2.pdf}
% \caption{The pinching-antenna system.}
% \label{sysmod}
% \end{figure}
We consider the extended system model to account for scenarios where the waveguide does not fully cover the diameter of the architecture. In such cases, the waveguide is constrained by the environment and is therefore shorter than the diameter of the circular region. The waveguide is deployed along a straight line parallel to the $x$-axis, with its starting and ending coordinates given by $(-l, 0, h)$ and $(l, 0, h)$, where the waveguide length is $2l (l < r)$.
The x-coordinate \(x_p\) of the pinching-antenna's placement is determined as
\begin{align}
x_p = 
\begin{cases}
    -l, & x_u < -l, \\
    x_u, & -l \leq x_u \leq l, \\
    l, & x_u > l.
\end{cases}
\end{align}
Such a placement strategy guarantees the minimal Euclidean distance between the pinching-antenna and the IoT device.
\subsection{partial coverage waveguide without propagation loss}
Let $D = \sqrt{(x_u - x_p)^2 + (y_u - y_p)^2}$ denote the horizontal distance between the PA and the IoT device. Then, we have
\begin{align}
D = 
\begin{cases}
    \sqrt{y_u^2+(x_u+l)^2}, & x_u < -l, \\
    y_u, & -l \leq x_u \leq l, \\
    \sqrt{y_u^2+(x_u-l)^2}, & x_u > l.
\end{cases}
\end{align}

In the case of a lossless dielectric waveguide, the SNR at the IoT device can be expressed as
\begin{align}
\gamma = \frac{\eta P_{t}}{\sigma^{2}(D^2 + h^2)}.
\label{gamma_with_D}
\end{align}

Based on (\ref{gamma_with_D}), the cumulative distribution function (CDF) of $D$ can be expressed as
\begin{align}
F_D(x) = \left\{
\begin{array}{ll}
1, & x \geq r, \\[2mm]
\frac{2\left(x \sqrt{r^{2}-x^2} + r^{2} \arcsin \left(\frac{x}{r}\right)\right)}{\pi r^{2}}, & \sqrt{r^2 - l^2} \leq x \leq r, \\[2mm]
\frac{2}{\pi r^2} \, \Theta(x), & r - l \leq x \leq \sqrt{r^2 - l^2}, \\[2mm]
\frac{4 x l + \pi x^2}{\pi r^2}, & 0 \leq x \leq r - l, \\[2mm]
0, & x \leq 0,
\end{array}
\right.
\end{align}
where $\Theta(x)$ can be expressed as
\begin{align}
\Theta(x) & \triangleq\ r^2 \cos^{-1} \left( \frac{l^2 + r^2 - x^2}{2 l r} \right) \\
& + x^2 \cos^{-1} \left( \frac{l^2 + x^2 - r^2}{2 l x} \right) - \pi l^2 + 4 x l \notag\\
& - \frac{1}{2} \sqrt{(-l + r + x)(l + r - x)(l - r + x)(l + r + x)}.
\end{align}
The outage probability can be expressed as
\begin{align}
    P_{\text {out }} &= \operatorname{Pr}\left(\gamma \leq \gamma_{\text {th}} \right) \notag\\
    &= \operatorname{Pr}\left(D^2 \geq \frac{\eta P_{t}e^{-\alpha(x_u+r)}}{\sigma^{2}\gamma_{\text {th}}} -h^{2}\right).
    \label{P_out_with_D}
\end{align}

Based on (\ref{P_out_with_D}) and the CDF of $D$, we have $P_{\text {out }} = 1-F_D(\sqrt{A})$. Therefore, the outage probability of the IoT device can be expressed as
\begin{align}
    P_{\text {out }}=\left\{\begin{array}{ll}0, & A \geq r^2, \\[3ex]
    1 - \Xi(A), & r^2-l^2 \leq A \leq r^2, \\[3ex]
    1 - \frac{2}{\pi r^2} \, \Theta(\sqrt{A}) , & (r-l)^2 \leq A \leq r^2-l^2, \\[3ex]
    1- \frac{4\sqrt{A}l+\pi A}{\pi r^2} , & 0 \leq A \leq (r-l)^2, \\[3ex]
    1, & A \leq 0,\end{array}\right.
\end{align}
where $\Xi(A) \triangleq \frac{2\left(\sqrt{A}\sqrt{r^{2}-A} + r^{2}\arcsin\!\left(\tfrac{\sqrt{A}}{r}\right)\right)}{\pi r^{2}}$. According to (\ref{gamma_with_D}), the average achievable rate is given by
\begin{align}
    R_{p}=\mathbb{E}\left[\log _{2}\left(1+\frac{\eta P_{t}}{\sigma^{2}(D^2 + h^2)}\right)\right].
\end{align}

\begin{lemma}
The closed-form expression for the average achievable rate is given by
\begin{align}
\bar{R}_p &\approx \frac{2}{\pi r^2 \ln 2} \bigg[
\frac{l \pi}{2V} \sum_{v=1}^{V} \sqrt{1 - \zeta_v^2}  f_1(\chi_v^{(1)})\notag \\
&\quad+
\frac{(r - l) \pi}{2V} \sum_{v=1}^{V} \sqrt{1 - \zeta_v^2}  f_2(\chi_v^{(2)})
\bigg],
\end{align}
where $\zeta_v = \cos\left( \frac{2v - 1}{2V} \pi \right), \chi_v^{(1)} = \frac{l}{2} \zeta_v + \frac{l}{2}, \chi_v^{(2)} = \frac{r - l}{2} \zeta_v + \frac{r + l}{2},$
\begin{align}
    f_1(y) &= 2 \sqrt{r^2 - y^2} \ln\left(r^2 - y^2 + h^2 + \frac{\eta P_t}{\sigma^2} \right) \notag \\
&\quad+ 4\sqrt{h^2 + \frac{\eta P_t}{\sigma^2}} \arctan\left( \frac{\sqrt{r^2 - y^2}}{\sqrt{h^2 + \frac{\eta P_t}{\sigma^2}}} \right) \notag\\
&\quad - 2 \sqrt{r^2 - y^2} \ln\left(r^2 - y^2 + h^2 \right) \notag \\
&\quad- 4h \arctan\left( \frac{\sqrt{r^2 - y^2}}{h} \right), \notag\\
f_2(y) &= 2 \sqrt{r^2 - y^2} \ln\left(r^2 + h^2 + l^2 - 2ly + \frac{\eta P_t}{\sigma^2} \right) \notag\\
&\quad + 4 \Lambda(y) \arctan\left( \frac{\sqrt{r^2 - y^2}}{\Lambda(y)} \right) \notag\\
&\quad - 2 \sqrt{r^2 - y^2} \ln\left(r^2 + h^2 + l^2 - 2ly \right) \notag \\
&\quad - 4\sqrt{h^2 + (y - l)^2} \arctan\left( \frac{\sqrt{r^2 - y^2}}{\sqrt{h^2 + (y - l)^2}} \right), \notag\\
\Lambda(y) &\triangleq \sqrt{h^2 + (y - l)^2 + \tfrac{\eta P_t}{\sigma^2}}.\notag
\end{align}

\end{lemma}
\begin{proof}
Please refer to Appendix C.
\end{proof}
\subsection{partial coverage waveguide with propagation loss}
When using a partial coverage waveguide with propagation loss, the SNR at the IoT device can be expressed as
\begin{align}
    \gamma =
    \begin{cases}
        \displaystyle\frac{\eta P_{t}}{\sigma^{2}(y_u^2 + h^2 + (x_u + l))^2}, & -r \leq x_u < -l, \\[3ex]
        \displaystyle\frac{\eta P_{t}e^{-\alpha(x_u+l)}}{\sigma^{2}(y_u^2 + h^2)}, & |x_u| \leq l, \\[3ex]
        \displaystyle\frac{\eta P_{t}e^{-2\alpha l}}{\sigma^{2}(y_u^2 + h^2 + (x_u - l)^2)}, & l < x_u \leq r.
    \end{cases}
\end{align}

By define
\begin{align}
    f(x_u)=\left\{\begin{array}{ll}C - h^{2} - (x_u + l)^2, & -r\leq x \leq -l, \\
    Ce^{-\alpha(x_u+l)} - h^{2}, & -l \leq x_u \leq l, \\ Ce^{-2\alpha l} - h^{2} - (x_u - l)^2, & l \leq x_u \leq r. \end{array}\right.
\end{align}
The outage probability can be given by
\begin{align}
    P_{\text {out }}
&= \operatorname{Pr}\left(\gamma \leq \gamma_{\text {thr }} \right) = \operatorname{Pr}\left(y_u \geq f(x_u)\right) \notag \\
&= \operatorname{Pr}\left(r^2 - x_u^2 \geq f(x_u)\right).
\end{align}
Let $D = \left\{\, (x_u, y_u)\;\middle|\; -r \leq x \leq r, \; f(x) \leq y^2 \leq r^2 - x_u^2 \, \right\}$, the outage probability can be expressed as
\begin{align}
    P_{\mathrm{out}} 
    = \iint\limits_{D}  \frac{1}{\pi r^2} dx_u\, dy_u.
    \label{P_out_integral_form}
\end{align}

Next, we first analyze the properties of the function with respect to the integration region $D$ in (\ref{P_out_integral_form}). 
\begin{lemma}
    The function $g(x_u) = r^2 - x_u^2 - f(x_u)$ attains a maximum in the interval $[-r, r]$.
\end{lemma}
\begin{proof}
   The first derivative of $g(x_u)$ with respect to $x_u$ in the interval $[-r, r]$ is given by
   \begin{align}
        \frac{dg(x_u)}{dx_u}=\left\{\begin{array}{ll}2l, & -r < x < -l, \\
        \alpha Ce^{-\alpha(x_u+l)} - 2x, & -l < x_u < l, \\- 2l, & l < x_u < r. \end{array}\right.
        \label{first_derivative}
   \end{align}
   
   Then the second derivative of $g(x_u)$ with respect to $x_u$ in the interval $[-r, r]$ can be expressed by
   \begin{align}
        \frac{d^2g(x_u)}{dx_u^2}=\left\{\begin{array}{ll}0, & -r < x < -l, \\
        -\alpha^2 Ce^{-\alpha(x_u+l)} - 2, & -l < x_u < l, \\0, & l < x_u < r. \end{array}\right.
        \label{second_derivative}
   \end{align}
   
According to (\ref{second_derivative}), the derivative \(\frac{dg(x_u)}{dx_u}\) is monotonically decreasing in the interval \(x_u \in [-l, \, l]\). Moreover, based on (\ref{first_derivative}), the derivative \(\frac{d g(x_u)}{d x_u} > 0\) for \(x_u \in [-r, \, 0]\), while the sign of the derivative at \(x_u = r\) cannot be determined in general. Therefore, $g(x_u)$ attains a maximum in the interval $[-r, r]$.
\end{proof}
According to Lemma 4, if  $\frac{dg(x_u)}{dx_u} = 0$ admits no solution with $g(l) \leq 0$, or only a single solution $x_u = d$ that satisfies $g(d) \leq 0$ within the interval $[-r, \, r]$, then $P_{\mathrm{out}} = 0$. In the following, we analyze the properties of the function $f(x_u)$.
\begin{lemma}
    The function $f(x_u)$ attains a maximum at $x_u = -l$.
\end{lemma}
\begin{proof}
    The first derivative of $f(x_u)$ with respect to $x_u$ in the interval $[-r, r]$ is given by
    \begin{align}
        \frac{df(x_u)}{dx_u}=\left\{\begin{array}{ll}-2(x_u + l), & -r < x < -l, \\
        -\alpha Ce^{-\alpha(x_u+l)}, & -l < x_u < l, \\- 2(x_u - l), & l < x_u < r. \end{array}\right.
        \label{first derivative of f(x_u)}
   \end{align}
   
   According to (\ref{first derivative of f(x_u)}), \(f(x_u)\) is monotonically increasing in the interval \([-r, \, -l]\) and monotonically decreasing in the intervals \([-l, \, l]\) and \([l, \, r]\). Therefore, the maximum of the function \(f(x_u)\) is attained at the point \(x_u = -l\).
\end{proof}
Based on Lemma 5, if $f(-l) < 0$, we have $P_{\mathrm{out}} = 1$. Let $\Omega(x_u) = Ce^{-\alpha(x_u+l)}$.
When \(g(x_u) = 0\) has two solutions, \(x_u = a\) and \(x_u = c\) with \(a < c\), the outage probability is given by
\begin{align}
    P_{\text{out}} &= 2\int_{a}^{c} \int_{\sqrt{\Omega(x_u)-h^2}}^{\rho(x_u)} \frac{1}{\pi r^2} \, dy_u \, dx_u \notag\\
    &= \frac{1}{\pi r^2} [
    r^2 \left( \arcsin\left( \frac{c}{r} \right) - \arcsin\left( \frac{a}{r} \right) \right) \notag \\
    &\quad + \left( c \rho(c) - a \rho(a) \right)
    ] \notag\\
    &\quad - \frac{4}{\alpha \pi r^2} \Bigg[
    h \arctan\left( \frac{\sqrt{\Omega(c) - h^2}}{h} \right) \notag \\
    &\quad - h\arctan\left( \frac{\sqrt{\Omega(a) - h^2}}{h} \right) \notag \\
    &\quad - \left(
    \sqrt{\Omega(c) - h^2}
    - \sqrt{\Omega(a) - h^2}
    \right)
    \Bigg].
\end{align}

When \(a\) lies within the interval \([-l, \, l]\) and \(c\) lies within the interval \([l, \, r]\), the outage probability is given by
\begin{align}
P_{\mathrm{out}} 
&= 2 \int_{a}^{l} \int_{\sqrt{\Omega(x_u) - h^2}}^{\rho(x_u)} \frac{1}{\pi r^2} \, dy_u dx_u \notag \\
&\quad+ 2 \int_{l}^{c} \int_{\sqrt{\Omega(l) - h^2 - (x_u - l)^2}}^{\rho(x_u)} \frac{1}{\pi r^2} \, dy_u dx_u \notag\\
&=\frac{2}{\pi r^2} \Bigg\{ \; \left[
\frac{1}{2} x_u \rho(x_u)
+ \frac{1}{2} r^2 \arcsin\left( \frac{x_u}{r} \right)
\right]_{a}^{c} \notag \\
&\quad - \Bigg[
\frac{1}{2} (c - l) \sqrt{\Omega(l) - h^2 - (c - l)^2} \notag\\
&\quad+ \frac{1}{2} \left(\Omega(l) - h^2\right) \arcsin\left( \frac{c - l}{\sqrt{\Omega(l) - h^2}} \right)
\Bigg]
\Bigg\} \notag \\
&\quad- \frac{4}{\alpha \pi r^2} \Bigg[
h \arctan\left( \frac{\sqrt{\Omega(l) - h^2}}{h} \right)\notag \\
&\quad- h\arctan\left( \frac{\sqrt{\Omega(a) - h^2}}{h} \right)\notag \\
&\quad- \left(
\sqrt{\Omega(l) - h^2}
- \sqrt{\Omega(a) - h^2}
\right)
\Bigg].
\end{align}

When \(a\) lies within the interval \([-r, \, -l]\) and \(c\) lies within the interval \([l, \, r]\), the outage probability is given by
\begin{align}
    P_{out} &= 2\int_{a}^{-l} \int_{\sqrt{C - h^2 - (x_u + l)^2}}^{\rho(x_u)} \frac{1}{\pi r^2} \, dy_u dx_u \notag \\
&\quad + 2\int_{-l}^{l} \int_{\sqrt{\Omega(x_u) - h^2}}^{\rho(x_u)} \frac{1}{\pi r^2} \, dy_u dx_u \notag \\
&\quad + 2\int_{l}^{c} \int_{\sqrt{\Omega(l) - h^2 - (x_u - l)^2}}^{\rho(x_u)} \frac{1}{\pi r^2} \, dy_u dx_u \notag\\
    &=\frac{2}{\pi r^2} \Bigg\{ \;
\left[
\frac{1}{2} x_u \rho(x_u) + \frac{1}{2} r^2 \arcsin\left(\frac{x_u}{r}\right)
\right]_{a}^{c} \notag \\
&\quad+ \Bigg[
\frac{1}{2} (a + l) \sqrt{C - h^2 - (a+ l)^2}\notag \\
&\quad+ \frac{1}{2} (C - h^2) \arcsin\left( \frac{a + l}{\sqrt{C - h^2}} \right)
\Bigg] \notag \\
&\quad - \Bigg[
\frac{1}{2} (c - l) \sqrt{\Omega(l) - h^2 - (c - l)^2} \notag\\
&\quad+ \frac{1}{2} \left(\Omega(l) - h^2\right) \arcsin\left( \frac{c - l}{\sqrt{\Omega(l) - h^2}} \right)
\Bigg]
\Bigg\} \notag \\
&\quad- \frac{4}{\alpha \pi r^2} \Bigg[
h \arctan\left( \frac{\sqrt{\Omega(l) - h^2}}{h} \right)\notag \\
&\quad- h\arctan\left( \frac{\sqrt{\Omega(-l) - h^2}}{h} \right)\notag \\
&\quad- \left(
\sqrt{\Omega(l) - h^2}
- \sqrt{\Omega(-l) - h^2}
\right)
\Bigg].
\end{align}

Next, we consider the case where  \(g(x_u) = 0\) has a single solution \(x_u = a\), and  \(f(x_u) = 0\) has a single solution \(x_u = b\). When \(a\) lies within the interval \([-r, \, -l]\) and \(b\) lies within the interval \([-l, \, l]\), the outage probability is given by
\begin{align}
    P_{\mathrm{out}} 
&= 2\int_{a}^{-l} \int_{\sqrt{C - h^2 - (x_u + l)^2}}^{\rho(x_u)} \frac{1}{\pi r^2}\, dy_u\, dx_u \notag\\
&\quad + 2\int_{-l}^{b} \int_{\sqrt{\Omega(x_u) - h^2}}^{\rho(x_u)} \frac{1}{\pi r^2}\, dy_u\, dx_u \notag\\ 
&\quad + 2\int_{b}^{r} \int_{0}^{\rho(x_u)} \frac{1}{\pi r^2}\, dy_u\, dx_u \notag \\
&=\frac{1}{2} + \frac{2}{\pi r^2} \Bigg[ \frac{C - h^2}{2} \arcsin\left( \frac{a + l}{\sqrt{C - h^2}} \right) \notag\\
&\quad +\frac{a + l}{2}\sqrt{C - h^2 - (a + l)^2} \notag \\
&\quad - \frac{a}{2}\sqrt{r^2 - a^2} - \frac{r^2}{2}\arcsin(\frac{a}{r}) \notag \\
&\quad - \frac{2}{\alpha} \left( \sqrt{C - h^2} - \sqrt{Ce^{-\alpha (l+b)} - h^2} \right)\notag \\
&\quad + \frac{2h}{\alpha} \Bigg(\arctan\left( \frac{\sqrt{C - h^2}}{h} \right) \notag \\
&\quad- \arctan\left( \frac{\sqrt{Ce^{-\alpha (l+b)} - h^2}}{h} \Bigg) \right) \Bigg].
\end{align}

When \(a\) lies within the interval \([-r, \, -l]\) and \(b\) lies within the interval \([l, \, r]\), the outage probability is given by
\begin{align}
    P_{\mathrm{out}} 
&= 2\int_{a}^{-l} \int_{\sqrt{C - h^2 - (x_u + l)^2}}^{\rho(x_u)} \frac{1}{\pi r^2} \, dy_u \, dx_u \notag\\
&\quad + 2\int_{-l}^{l} \int_{\sqrt{\Omega(x_u) - h^2}}^{\rho(x_u)} \frac{1}{\pi r^2} \, dy_u \, dx_u \notag\\
&\quad + 2\int_{l}^{b} \int_{\sqrt{\Omega(l) - h^2 - (x_u - l)^2}}^{\rho(x_u)} \frac{1}{\pi r^2} \, dy_u \, dx_u \notag\\
&\quad + 2\int_{b}^{r} \int_{0}^{\rho(x_u)} \frac{1}{\pi r^2} \, dy_u \, dx_u \notag \\
&=\frac{2}{\pi r^2} \Bigg\{\left[
\frac{1}{2} x_u \rho(x_u)
+ \frac{1}{2} r^2 \arcsin\left(\frac{x_u}{r}\right)
\right]_{a}^{r} \notag\\
&\quad + \Bigg[
    \frac{1}{2} (a + l) \sqrt{C-h^2-(a+l)^2} \notag \\
    &\quad+ \frac{1}{2} (C-h^2) \arcsin\left(\frac{a+l}{\sqrt{C-h^2}}\right)
\Bigg] \notag\\[1.5ex]
&\quad - \Bigg[
    \frac{1}{2} (b - l) \sqrt{\Omega(l)-h^2-(b-l)^2} \notag \\
    &\quad+ \frac{1}{2} (\Omega(l)-h^2) \arcsin\left(\frac{b-l}{\sqrt{\Omega(l)-h^2}}\right)
\Bigg]
\Bigg\} \notag \\
&\quad- \frac{4}{\alpha \pi r^2} \Bigg[
h \arctan\left( \frac{\sqrt{\Omega(l) - h^2}}{h} \right)\notag \\
&\quad- h\arctan\left( \frac{\sqrt{\Omega(-l) - h^2}}{h} \right)\notag \\
&\quad- \left(
\sqrt{\Omega(l) - h^2}
- \sqrt{\Omega(-l) - h^2}
\right)
\Bigg].
\end{align}

When \(a\) lies within the interval \([-l, \, l]\) and \(b\) also lies within the interval \([-l, \, l]\), the outage probability is given by
\begin{align}
P_{out} & = 2\int_{a}^{b} \int_{\sqrt{\Omega(x_u)-h^2}}^{\rho(x_u)} \frac{1}{\pi r^2}dy_udx_u \notag\\ 
&\quad+ 2\int_{b}^{r} \int_{0}^{\rho(x_u)} \frac{1}{\pi r^2}dy_udx_u \notag\\
&= \frac{1}{\pi r^2} \left[
r^2 \left( \frac{\pi}{2} - \arcsin\left( \frac{a}{r} \right) \right) - a \rho(a)
\right]  \notag\\
&\quad - \frac{4}{\alpha \pi r^2} \Bigg[
h\arctan\left( \frac{\sqrt{\Omega(b) - h^2}}{h} \right)\notag\\
&\quad- h\arctan\left( \frac{\sqrt{\Omega(a) - h^2}}{h} \right) \notag\\
&\quad-\sqrt{\Omega(b) - h^2}
+ \sqrt{\Omega(a) - h^2}
\Bigg].
\end{align}

When \(a\) lies within the interval \([-l, \, l]\) and \(b\) lies within the interval \([l, \, r]\), the outage probability is given by
\begin{align}
    P_{\mathrm{out}} &= 2\int_{a}^{l} \int_{\sqrt{\Omega(x_u)-h^2}}^{\sqrt{r^2 - x_u^2}} \frac{1}{\pi r^2}dy_udx_u \notag\\
    &\quad+ 2\int_{l}^{b} \int_{\sqrt{\Omega(l)-h^2- (x_u - l)^2}}^{\sqrt{r^2 - x_u^2}} \frac{1}{\pi r^2}dy_udx_u \notag\\
    &\quad+ 2\int_{b}^{r} \int_{0}^{\sqrt{r^2 - x_u^2}} \frac{1}{\pi r^2}dy_udx_u \notag\\
    &=\frac{2}{\pi r^2} \Bigg\{
\left[
\frac{1}{2} x_u\, \rho(x_u)
+ \frac{1}{2} r^2\, \arcsin\left( \frac{x_u}{r} \right)
\right]_{a}^{r} \notag \\[1ex]
&\quad - \left[
\frac{1}{2} (b - l)\, \sqrt{ \Omega(l) - h^2 - (b - l)^2 }
\right. \notag \\
&\quad + \left.
\frac{1}{2} \left( \Omega(l) - h^2 \right)
\arcsin\left( \frac{b - l}{ \sqrt{ \Omega(l) - h^2 } } \right)
\right]
\Bigg\} \notag \\[1ex]
&\quad - \frac{4}{\alpha \pi r^2} \Bigg[
h\, \arctan\left( \frac{ \sqrt{ \Omega(l) - h^2 } }{ h } \right) \notag \\
&\quad- h\, \arctan\left( \frac{ \sqrt{ \Omega(a) - h^2 } }{ h } \right) \notag \\
&\quad - \left( \sqrt{ \Omega(l) - h^2 }
- \sqrt{ \Omega(a) - h^2 } \right)
\Bigg].
\end{align}

Finally, we consider the case where  \(f(x) = 0\) has two solutions $x_u = a$ and $x_u = b$ with $a<b$. When \(a\) lies within the interval \([-r, \, -l]\) and \(b\) lies within the interval \([-l, \, l]\), the outage probability is given by
\begin{align}
    P_{\mathrm{out}} &= 2\int_{-r}^{a} \int_{0}^{\sqrt{r^2 - x_u^2}} \frac{1}{\pi r^2}dy_udx_u \notag\\
    &\quad+2\int_{a}^{-l} \int_{\sqrt{C-h^2 - (x_u+l)^2}}^{\sqrt{r^2 - x_u^2}} \frac{1}{\pi r^2}dy_udx_u \notag\\
    &\quad+2\int_{-l}^{b} \int_0^{\sqrt{r^2 - x_u^2}} \frac{1}{\pi r^2}dy_udx_u \notag\\
    &\quad+ 2\int_{b}^{l} \int_{\sqrt{Ce^{-2 \alpha l}-h^2- (x_u - l)^2}}^{\sqrt{r^2 - x_u^2}} \frac{1}{\pi r^2}dy_udx_u \notag\\
    &\quad+ 2\int_{l}^{r} \int_{0}^{\sqrt{r^2 - x_u^2}} \frac{1}{\pi r^2}dy_udx_u \notag\\
    & =1 + \frac{2}{\pi r^2} \Bigg[ \frac{C - h^2}{2} \arcsin\left( \frac{a + l}{\sqrt{C - h^2}} \right) \notag\\
&\quad - \frac{2}{\alpha} \left( \sqrt{C - h^2} - \sqrt{Ce^{-\alpha (l+b)} - h^2} \right)\notag \\
&\quad + \frac{2h}{\alpha} \Bigg(\arctan\left( \frac{\sqrt{C - h^2}}{h} \right) \notag \\
&\quad- \arctan\left( \frac{\sqrt{Ce^{-\alpha (l+b)} - h^2}}{h} \Bigg) \right) \Bigg].
\end{align}

When \(a\) lies within the interval \([-r, \, -l]\) and \(b\) lies within the interval \([l, \, r]\), the outage probability is given by
\begin{align}
    P_{\mathrm{out}} &= 2\int_{-r}^{a} \int_{0}^{\sqrt{r^2 - x_u^2}} \frac{1}{\pi r^2}dy_udx_u \notag\\
    &\quad+2\int_{a}^{-l} \int_{\sqrt{C-h^2 - (x_u + l)^2}}^{\sqrt{r^2 - x_u^2}} \frac{1}{\pi r^2}dy_udx_u \notag\\
    &\quad+2\int_{-l}^{l} \int_{\sqrt{Ce^{-\alpha(x_u+l)}-h^2}}^{\sqrt{r^2 - x_u^2}} \frac{1}{\pi r^2}dy_udx_u \notag\\
    &\quad+ 2\int_{l}^{b} \int_{\sqrt{Ce^{-2 \alpha l}-h^2- (x_u - l)^2}}^{\sqrt{r^2 - x_u^2}} \frac{1}{\pi r^2}dy_udx_u \notag\\
    &\quad+ 2\int_{b}^{r} \int_{0}^{\sqrt{r^2 - x_u^2}} \frac{1}{\pi r^2}dy_udx_u \notag\\
    &=  1 + \frac{2}{\pi r^2} \Bigg[ \frac{C - h^2}{2} \arcsin\left( \frac{a + l}{\sqrt{C - h^2}} \right) \\
&\quad - \frac{Ce^{-2\alpha l} - h^2}{2} \arcsin\left( \frac{b - l}{\sqrt{Ce^{-2\alpha l} - h^2}} \right) \notag\\
&\quad - \frac{2}{\alpha} \left( \sqrt{C - h^2} - \sqrt{Ce^{-2\alpha l} - h^2} \right)\notag \\
&\quad + \frac{2h}{\alpha} \Bigg(\arctan\left( \frac{\sqrt{C - h^2}}{h} \right) \notag \\
&\quad- \arctan\left( \frac{\sqrt{Ce^{-2\alpha l} - h^2}}{h} \Bigg) \right) \Bigg].
\end{align}
\begin{lemma}
The average achievable rate for a partial coverage waveguide with propagation loss can be expressed as
    \begin{align}
    R_p \approx \frac{1}{n r^2 \ln 2} \Bigg[
    & 2 l \sum_{k=1}^n F_1(x_{1, k}) + (r-l) \sum_{k=1}^n F_2(x_{2, k}) \notag \\
    & + (r-l) \sum_{k=1}^n F_3(x_{3, k})
    \Bigg].   \label{average_achievable_rate_with_waveguide_propagation_loss}
    \end{align}
where $n$ is the number of quadrature points, and
$ x_{1, k} = l t_k, x_{2, k} = \frac{r-l}{2} t_k + \frac{r+l}{2}, x_{3, k} = \frac{r-l}{2}t_k - \frac{r+l}{2}, t_k = \cos ( \frac{2k-1}{2n} \pi ), k=1, 2, \dots, n.
$
The integrand functions in (\ref{average_achievable_rate_with_waveguide_propagation_loss}) for each segment are given by
\begin{align*}
F_1(x_u) &= 
\sqrt{r^2 - x_u^2} \ln\left(1 + \frac{\tfrac{\eta P_t}{\sigma^2} e^{-\alpha(x_u+l)}}{r^2 - x_u^2 + h^2}\right) \\
&\quad + 2\sqrt{\Lambda_1(x_u)} \arctan\left( \frac{\sqrt{r^2 - x_u^2}}{\sqrt{\Lambda_1(x_u)}} \right) \\
&\quad - 2h \arctan\left( \frac{\sqrt{r^2 - x_u^2}}{h} \right), \\[2ex]
F_2(x_u) &= 
\sqrt{r^2 - x_u^2} \ln\left(1 + \frac{\tfrac{\eta P_t}{\sigma^2} e^{-2 \alpha l}}{r^2 - x_u^2 + \Lambda_2(x_u)} \right) \\
&\quad + 2\sqrt{\Lambda_2(x_u)} \arctan\left( \frac{ \sqrt{ r^2 - x_u^2 } }{ \sqrt{\Lambda_2(x_u)} } \right) \\
&\quad - 2\sqrt{ h^2 + (x_u - l)^2 } \arctan\left( \frac{ \sqrt{ r^2 - x_u^2 } }{ \sqrt{ h^2 + (x_u - l)^2 } } \right), \\[2ex]
F_3(x_u) &= 
\sqrt{ r^2 - x_u^2 } \ln\left( 1 + \frac{ \tfrac{ \eta P_t }{ \sigma^2 } }{ r^2 - x_u^2 + \Lambda_3(x_u) } \right) \\
&\quad + 2\sqrt{\Lambda_3(x_u)} \arctan\left( \frac{ \sqrt{ r^2 - x_u^2 } }{ \sqrt{\Lambda_3(x_u)} } \right) \\
&\quad - 2\sqrt{ h^2 + (x_u + l)^2 } \arctan\left( \frac{ \sqrt{ r^2 - x_u^2 } }{ \sqrt{ h^2 + (x_u + l)^2 } } \right), \\
\Lambda_1(x_u) &\triangleq h^2 + \frac{\eta P_t}{\sigma^2} e^{-\alpha(x_u+l)}, \\[1ex]
\Lambda_2(x_u) &\triangleq h^2 + (x_u - l)^2 + \frac{\eta P_t}{\sigma^2} e^{-2\alpha l}, \\[1ex]
\Lambda_3(x_u) &\triangleq h^2 + (x_u + l)^2 + \frac{\eta P_t}{\sigma^2}.
\end{align*}
\end{lemma}
\begin{proof}
Please refer to Appendix D.
\end{proof}

% \begin{table}[t]
% 	\begin{center}
% 		\caption{Table of Simulation Parameters.}
% 		\label{table1}
% 		\renewcommand\arraystretch{1.5}
% 		\begin{tabular}{| c | c |}
% 			\hline
% 			Intensity $\lambda$ (IoT devices/m$^2$) & 0.1 \cite{LJB}\\
% 			\hline
% 			Power splitting factor $\rho$ & 0.6\\
% 			\hline
% 			Downlink time factor $\beta$ & 0.6\\
% 			\hline
% 			Path-loss exponent $\alpha_1$ for the Rayleigh channel & 2.8\\
% 			\hline
% 			Path-loss exponent $\alpha_2$ for the Rician channel & 2.3\\
% 			\hline
% 			Vertical height of the RIS $H$ (m)& 5\\
% 			\hline
% 			The radius of the circular target area $L$ (m)& 5\\
% 			\hline
% 			The distance from the BS to the RIS $d_\mathrm{SR} (m)$ & 10\\ 
% 			\hline
% 			Decoding power consumption coefficient $\varpi$ & $10^{-10}$ \cite{AM2} \\
% 			\hline
% 			Energy conversion efficiency of IoT device $\eta$ & 0.85\\ 
% 			\hline
% 			Noise power at the IoT device $\sigma_{\mathrm{{dl}}, k}^2$ (dBm) & -74\\
% 			\hline
% 			Noise power at the BS $\sigma_{\mathrm{up}}^2$ (dBm) & -74\\
% 			\hline
% 			Transmission duration $\tau_c$ (sec) & 1\\
% 			\hline 
% 			Uplink transmission rate threshold $C_\mathrm{up, th}$ (bits) & 0.1 \\
% 			\hline 
% 		\end{tabular}
% 	\end{center}
% \end{table}

\section{Numerical results}
In this section, we comprehensively evaluate the outage probability and the average achievable rate under various waveguide configurations to validate the analytical results. Unless otherwise specified, the key system parameters are configured as follows: the noise power is set to $\sigma^2 = -90\, \mathrm{dBm}$, the carrier frequency is $f_c = 28\, \mathrm{GHz}$, and the antenna height is fixed at $h = 10\, \mathrm{m}$. The effective IoT device region is modeled as a disk with radius $r = 25\, \mathrm{m}$, and the outage threshold for the SNR is set to $\gamma_{\mathrm{th}} = 100$. The waveguide attenuation coefficient is $\alpha = 0.02$, and the speed of light is $c = 3 \times 10^8\, \mathrm{m/s}$.

\begin{figure}[!t]
    \centering
    \includegraphics[width=0.9\linewidth]{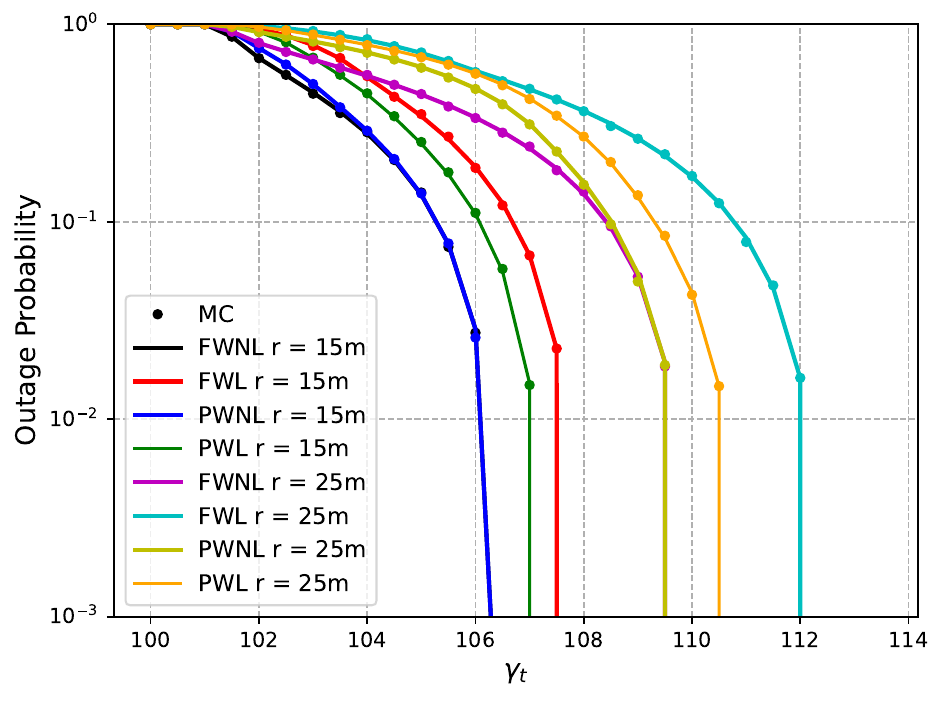}
    \caption{Outage probability versus transmit SNR under different waveguide length.}
    \label{fig:outage_snr_r}
\end{figure}
Fig.~\ref{fig:outage_snr_r} depicts the outage probability versus transmit SNR $\gamma_t = \frac{P_t}{\sigma^2}$ for different waveguide configurations and coverage radii. Both Monte-Carlo (MC) simulation results and closed-form analytical expressions are included for comparison, covering four representative scenarios: full coverage waveguide experiences no propagation loss (FWNL), full coverage waveguide with propagation loss (FWL), partial coverage waveguide experiences no propagation loss (PWNL), and partial coverage waveguide with propagation loss (PWL). As observed, the analytical results match well with the MC simulations, thereby validating the accuracy of the derived expressions. In all considered scenarios, the outage probability decreases rapidly as SNR increases, indicating significant improvements in link reliability at higher SNR. Comparing different configurations, it is evident that both waveguide propagation loss and partial coverage significantly deteriorate outage performance. In particular, scenarios with waveguide propagation loss (FWL and PWL) present a clear outage probability floor due to additional propagation attenuation. Furthermore, a larger coverage radius leads to higher outage probability in all cases, as IoT device are distributed over a wider area, resulting in greater average path loss. It can be observed that FWNL consistently attains the lowest outage probability, while FWL exhibits the poorest performance. This underscores the critical importance of both minimizing waveguide propagation loss and maintaining continuous coverage to ensure robust information transfer. These results quantitatively demonstrate the joint impact of waveguide deployment strategy, propagation loss, and coverage area on system outage performance.

\begin{figure}[!t]
    \centering
    \includegraphics[width=0.9\linewidth]{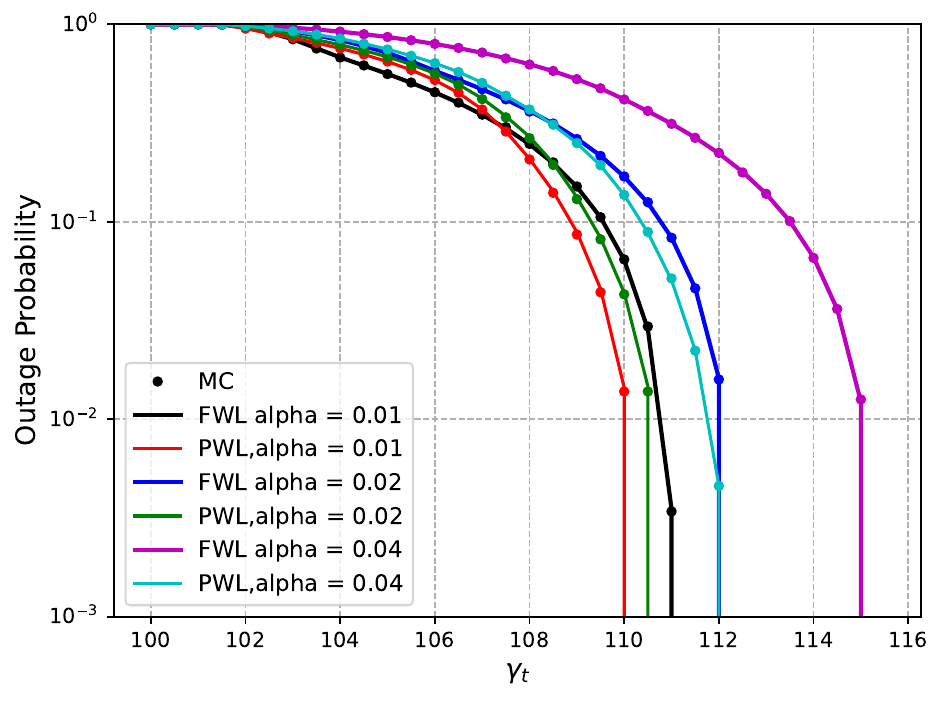}
    \caption{Outage probability versus transmit SNR under different waveguide attenuation coefficients.}
    \label{fig:outage_snr_alpha}
\end{figure}
Fig.~\ref{fig:outage_snr_alpha} illustrates the effect of the waveguide attenuation coefficient $\alpha$ on the outage probability for both FWL and PWL configurations. As $\alpha$ increases from $0.01$ to $0.04$, the outage probability rises notably in both configurations, reflecting the more severe propagation loss introduced by a larger attenuation coefficient. Notably, for $\alpha = 0.02$ and $0.04$, when the pinching-antenna is located closest to the IoT device, the FWL case consistently exhibits a higher outage probability than the PWL case. This is attributed to the excessive waveguide propagation loss under full coverage, which outweighs the potential benefits of continuous waveguide propagation. The gap between FWL and PWL widens as $\alpha$ increases, highlighting the pronounced impact of waveguide propagation loss on system reliability. The analytical and MC results show excellent agreement, thereby verifying the validity of the analysis. These indicate that controlling waveguide attenuation is essential to maintain low outage probability, particularly when partial coverage waveguide deployment is considered.

\begin{figure}[!t]
    \centering
    \includegraphics[width=0.9\linewidth]{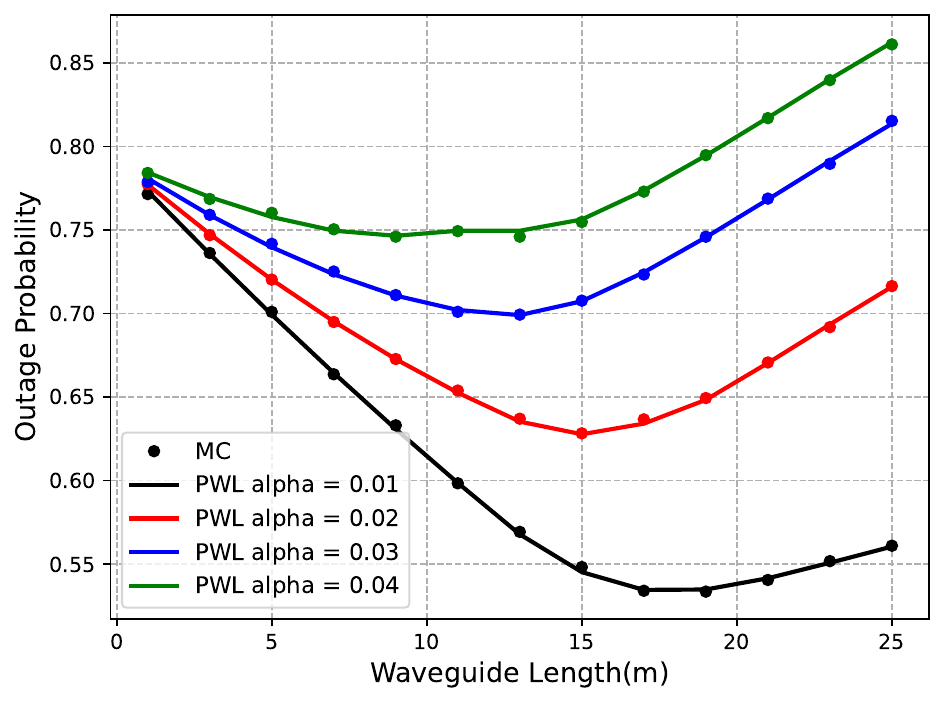}
    \caption{Outage probability versus waveguide length under different waveguide configurations.}
    \label{fig:outage_length}
\end{figure}

\begin{figure}[!t]
    \centering
    \includegraphics[width=0.9\linewidth]{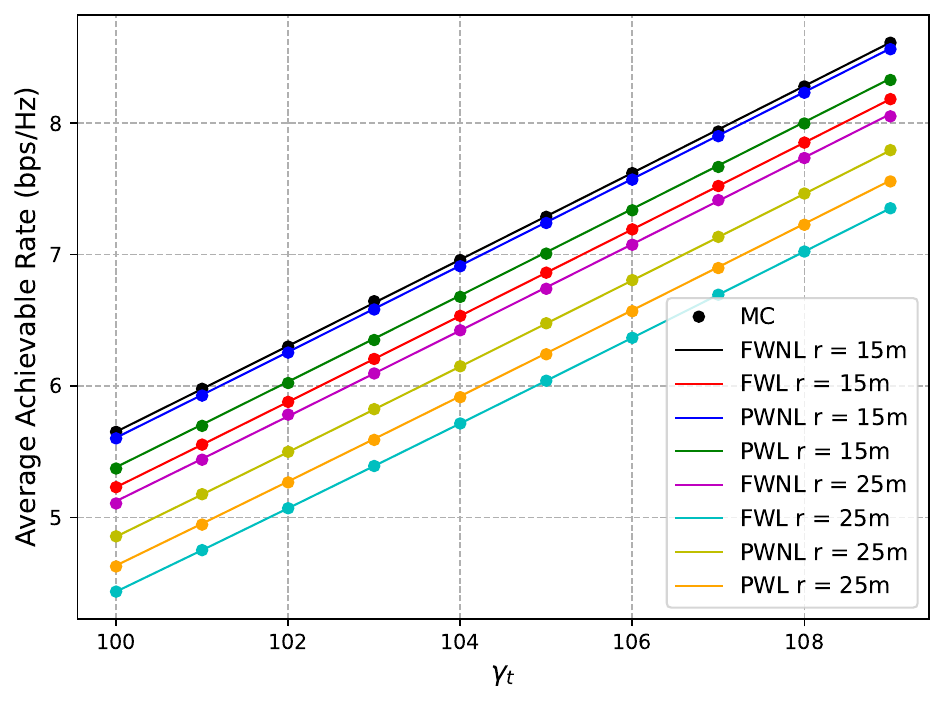}
    \caption{Average achievable rate versus SNR under different waveguide length.}
    \label{fig:rate_r}
\end{figure}

\begin{figure}[!t]
    \centering
    \includegraphics[width=0.9\linewidth]{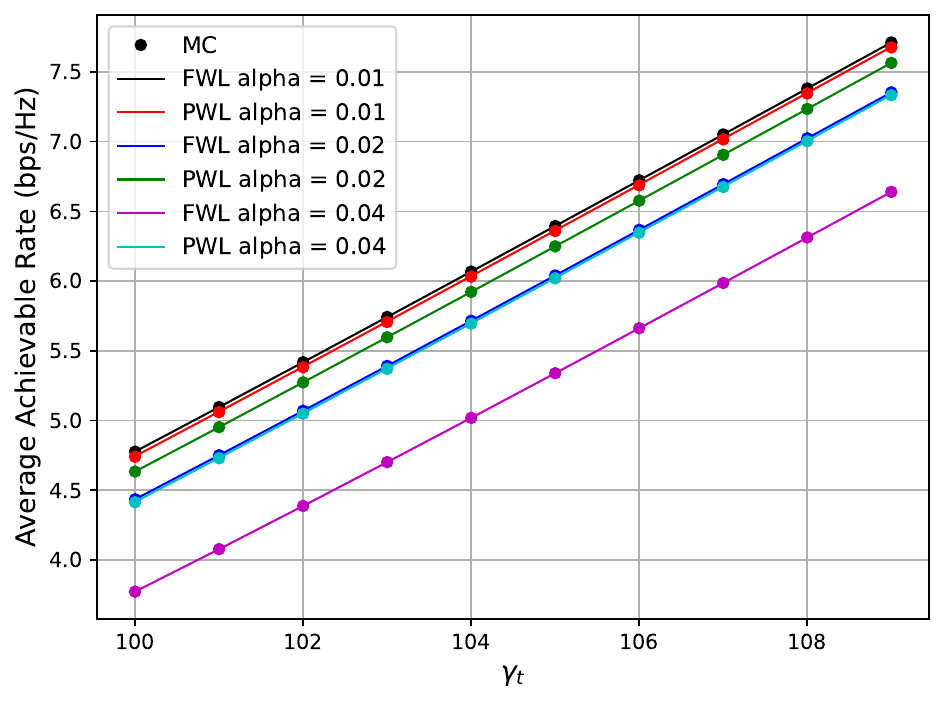}
    \caption{Average achievable rate versus SNR under different waveguide attenuation coefficients.}
    \label{fig:rate_alpha}
\end{figure}

\begin{figure}[!t]
	\centering
	\includegraphics[width=0.9\linewidth]{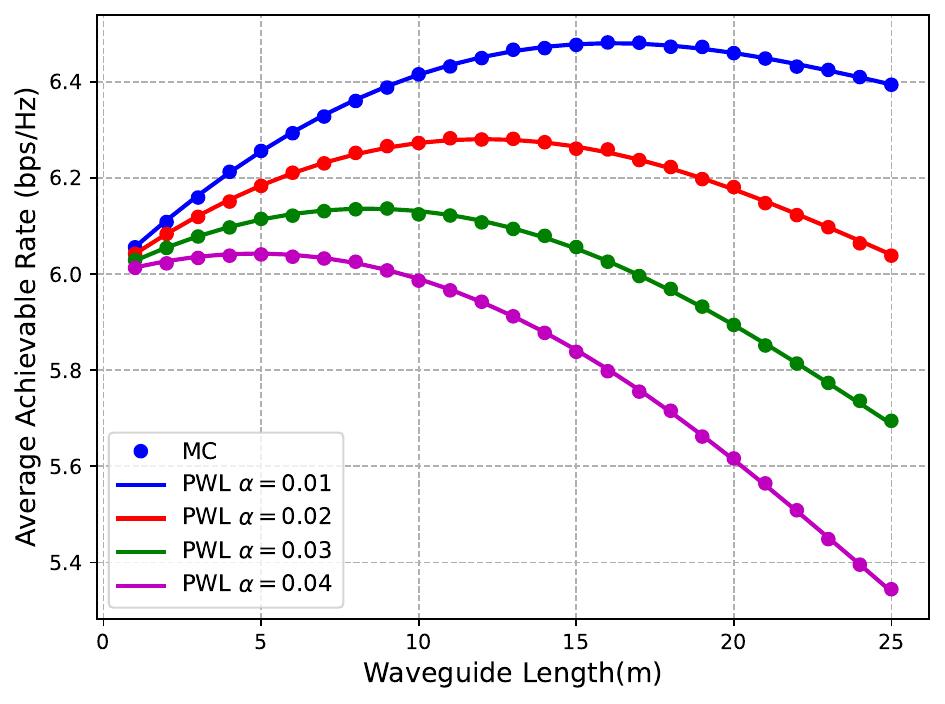}
	\caption{Average achievable rate versus waveguide length under different waveguide attenuation coefficients.}
	\label{fig:rate_length}
\end{figure}

Fig.~\ref{fig:outage_length} presents the outage probability versus waveguide length $l$ for different values of the attenuation coefficient $\alpha$ in the PWL scenario, with $\gamma_t=105\, \mathrm{dB}$.  A distinct non-monotonic trend is observed. Specifically, as $l$ increases from a small value, the outage probability initially decreases due to more IoT device benefiting from guided transmission. However, beyond a certain length, the accumulated waveguide propagation loss causes the outage probability to increase. Higher values of $\alpha$ not only elevate the minimum achievable outage probability but also reduce the optimal waveguide length that minimizes outage. These results reveal a crucial design trade-off, where extending the waveguide can improve coverage, but excessive length under large attenuation can be detrimental. Therefore, optimizing waveguide length according to specific attenuation characteristics is necessary for robust and reliable transmission.

Fig.~\ref{fig:rate_r} shows the average achievable rate versus SNR for different coverage radii with $r = 15\, \mathrm{m}$ and $r = 25\, \mathrm{m}$. It can be seen that, for a given SNR, the FWNL scenario provides the highest rate, while PWL yields the lowest. Increasing the coverage radius leads to reduced achievable rates due to the larger average IoT device distance and increased path loss. The introduction of waveguide propagation loss results in noticeable performance degradation, especially for larger radii. Additionally, partial coverage waveguide further diminishes the achievable rate, and this effect is magnified when combined with propagation loss. These results highlight the significant impact of both waveguide propagation loss and coverage strategy on system throughput.

Fig.~\ref{fig:rate_alpha} presents the average achievable rate versus SNR for different waveguide attenuation coefficients $\alpha$ $(0.01, 0.02, 0.04)$ under both FWL and PWL scenarios. For a given SNR, the achievable rate decreases monotonically with increasing $\alpha$ for both configurations. The degradation is more pronounced in the FWL scenario, highlighting the compounded impact of waveguide attenuation when the entire diameter is covered. For each $\alpha$, the PWL scenario consistently outperforms FWL, emphasizing the role of coverage strategy in mitigating the negative effects of waveguide attenuation. The performance gap between PWL and FWL widens with increasing $\alpha$, indicating that partial coverage becomes relatively more favorable in high-loss regimes. These underscore the importance of minimizing waveguide attenuation to maximize system throughput in practical deployments.

Fig.~\ref{fig:rate_length} illustrates the average achievable rate versus waveguide length $L$ for various attenuation coefficients $\alpha$ ($0.01, 0.02, 0.03, 0.04$) under the PWL scenario. The results reveal a non-monotonic trend, where the average achievable rate increases with $L$, reaches a maximum, and then decreases as $L$ continues to grow. This indicates the existence of an optimal waveguide length that balances reduced free-space loss and accumulated waveguide attenuation. Higher $\alpha$ values decrease both the maximum point of the rate curve and the associated optimal $L$, which emphasizes the negative effect of increased loss on coverage extension. 
%These emphasize the need to carefully balance waveguide length and attenuation in system design to optimize throughput in PASS.

\section{Conclusion}
This paper has presented the first analytical investigation of PASS in circular indoor environments, developing a unified geometric–propagation framework that jointly incorporates PA placement, IoT device distribution, and waveguide attenuation. Closed-form expressions for outage probability and average achievable rate were derived and validated through MC simulations under four representative configurations. In the full coverage waveguide without propagation loss case, PASS achieves the performance upper bound, consistently yielding the lowest outage probability and highest average rate. When propagation loss is included, however, full coverage waveguide suffers from severe attenuation, leading to an outage floor and reduced rate performance. For partial coverage waveguide without propagation loss, the analysis revealed the benefits of dynamic antenna placement under geometric constraints, whereas the combination of partial coverage and propagation loss demonstrated a distinct non-monotonic dependence of performance on waveguide length, with the optimal length decreasing as the attenuation coefficient increases. 
%Collectively, these results highlight the critical trade-off between coverage extension and attenuation, offering practical guidelines for optimizing waveguide deployment in PASS.
%For future work, we aim to maximize the average uplink data size by jointly optimizing the transmit beamforming of the BS, the phase shift of the RIS, the proportion coefficient of PS and the time allocation factor on the premise of meeting the average downlink data size requirements for the $k$-th IoT device.

\appendices
\setcounter{equation}{0}
\renewcommand\theequation{A.\arabic{equation}}

\section{Proof of Lemma 1}\label{proofone}
When the waveguide propagation loss is negligible, the average achievable rate can be derived as.
\begin{align}
R_{p} 
&= \frac{1}{\pi r^2 \ln 2} \int_{-r}^{r} \int_{-\rho(x_u)}^{\rho(x_u)} 
    \ln \left(1 + \frac{\eta P_t}{\sigma^{2}(y_u^2 + h^2)}\right) dx_u dy_u \notag \\
&= \frac{4}{\pi r^2 \ln 2} \int_0^r \rho(x_u) 
    \ln \left(1 + \frac{\eta P_t}{\sigma^{2}(y_u^2 + h^2)}\right) dy_u \notag \\
&= \frac{4}{\pi r^2 \ln 2} \int_0^r \rho(x_u) 
    \Big[\ln \left(\eta P_t + \sigma^2(y_u^2 + h^2)\right) \notag \\
    &\quad- \ln \left(\sigma^2(y_u^2 + h^2)\right)\Big] dy_u \notag \\
&= \frac{4}{\pi r^2 \ln 2} r^2 \Bigg(
    \int_0^1 \sqrt{1 - t^2} \ln(\eta P_t + \sigma^2 h^2) dt \notag\\
& \quad + \int_0^1 \sqrt{1 - t^2} \ln\left(1 + \frac{\sigma^2 r^2 t^2}{\eta P_t + \sigma^2 h^2}\right) dt         \notag \\
&\quad - \int_0^1 \sqrt{1 - t^2} \left(\ln(\sigma^2 h^2) + \ln(1 + \frac{r^2 t^2}{h^2})\right) dt
\Bigg).
\end{align}

Based on \cite[(4.295.27)]{edition2007table}, the average achievable rate can be derived as
\begin{align}
R_{p} 
&= \frac{4}{\pi r^2 \ln 2} \Bigg(
    \frac{\pi r^2}{4} \ln(B) 
    + \frac{\pi r^2}{2} \ln\left(\frac{1 + \Gamma}{2}\right) 
    + \frac{\pi r^2}{4}  \frac{1 - \Gamma}{1 + \Gamma}\notag \\
    &\quad - \frac{\pi r^2}{4} \ln(\sigma^2 h^2) 
    - \frac{\pi r^2}{2} \left[
        \ln\left(\frac{1 + \Lambda}{2}\right)
        + \frac{1}{2}  \frac{1 - \Lambda}{1 + \Lambda}
    \right]
\Bigg) \notag \\
&= \frac{1}{\ln 2} \ln\left(1 + \frac{\eta P_t}{\sigma^2 h^2}\right) + \frac{2}{\ln 2} \Bigg[
    \ln\left(\frac{1 + \Gamma}{1 + \Lambda}\right) 
    + \frac{1}{2}  \frac{1 - \Gamma}{1 + \Gamma} \notag \\
    &\qquad \quad\quad\qquad\quad\quad\qquad\qquad\qquad - \frac{1}{2}  \frac{1 - \Lambda}{1 + \Lambda}
\Bigg].
\end{align}

\setcounter{equation}{0}
\renewcommand\theequation{\textsc{B}.\arabic{equation}}
\section{Proof of Lemma 2}\label{prooftwo}
  Considering the waveguide propagation loss, the outage probability and achievable rate can be derived as
\begin{align}
R_p 
&= \frac{1}{\pi r^2 \ln 2} \int_{-r}^{r} \int_{-\rho(y_u)}^{\rho(y_u)} 
\ln\left( 1 + \frac{\Omega(x_u)}{\Psi(y_u)} \right) dx_u dy_u \notag \\[2mm]
&= \frac{1}{\pi r^2 \ln 2} 
\bigg[
\int_{-r}^{r} \int_{-\rho(y_u)}^{\rho(y_u)} \ln\left( \Omega(x_u) + \Psi(y_u) \right) dx_u dy_u\notag \\
&\quad- \int_{-r}^{r} \int_{-\rho(y_u)}^{\rho(y_u)} \ln\left( \Psi(y_u) \right) dx_u dy_u 
\bigg] \notag \\[2mm]
&= \frac{1}{\pi r^2 \ln 2} \int_{-r}^{r} \int_{-\rho(y_u)}^{\rho(y_u)} 
\ln\left( \Omega(x_u) + \Psi(y_u) \right) dx_u dy_u \notag \\
&\quad - \frac{1}{\pi r^2 \ln 2}
\left[
\int_{-r}^{r} 2 \rho(y_u) \ln\left( \Psi(y_u) \right) dy_u
\right] \notag \\
& =\frac{1}{\pi r^2 \ln 2} \Bigg[
\int_{-r}^{r} \int_{-\rho(y_u)}^{\rho(y_u)} 
\ln \left( \Omega(x_u) + \Psi(y_u) \right) dx_u dy_u \notag \\
&\quad - \frac{\pi r^2}{2} \ln(\Psi(0)) - \pi r^2 \left\{
\ln\left(\frac{1 + \Lambda}{2}\right)
+ \frac{1}{2}  \frac{1 - \Lambda}{1 + \Lambda}
\right\}
\Bigg].
\label{R_p}
\end{align}

The integral term in the first component of expression (\ref{R_p}) can be further derived using \cite[(2.728.2)]{edition2007table} as
\begin{align}
&\int_{-r}^{r} \int_{-\rho(y_u)}^{\rho(y_u)} 
\ln \left( \Omega(x_u) + \Psi(y_u) \right) dx_u dy_u\notag\\
&= \frac{1}{\alpha} \int_{-r}^{r}  \int_{v_2}^{v_1} 
\frac{\ln \left( \eta P_t v + \Psi(y_u) \right)}{v} \, dv \, dy_u \notag\\
&= \int_{-r}^{r} \bigg[
2 \ln \left( \Psi(y_u) \right)  \rho(y_u)
+ \frac{1}{\alpha} \mathrm{Li}_2\left( -\frac{\eta P_t}{\Psi(y_u)} \right)\notag \\
&\qquad \qquad \qquad \qquad - \frac{1}{\alpha} \mathrm{Li}_2\left( -\frac{\eta P_t}{\Psi(y_u)} \right)
\bigg] dy_u \notag\\
&= \pi r^2 \ln(\sigma^2 h^2)
+ 2\pi r^2 \left\{
\ln\left(\frac{1 + \Lambda}{2}\right)
+ \frac{1}{2}  \frac{1 - \Lambda}{1 + \Lambda}
\right\} \notag\\
&\quad - \int_{-r}^{r} \left[
\frac{1}{\alpha} \, \mathrm{Li}_2\left( -\frac{\eta P_t}{\Psi(y_u)} \right)
+ \frac{1}{\alpha} \, \mathrm{Li}_2\left( -\frac{\eta P_t}{\Psi(y_u)} \right)
\right] dy_u.
\end{align}

Then applying the Gauss–Chebyshev quadrature method\cite{kumar2022performance}, a closed-form approximation of the average achievable rate can be expressed as
\begin{align}
R_{p} 
&= \pi r^2 \ln(\sigma^2 h^2) + 2\pi r^2 \left\{ \ln\left(\frac{1 + \Lambda}{2}\right)
+ \frac{1}{2}  \frac{1 - \Lambda}{1 + \Lambda} \right\}\notag \\
&\quad+ \frac{r\pi}{V} \sum_{m=1}^{V} \sqrt{1 - x_m^2} \Bigg[
    \frac{1}{\alpha} \mathrm{Li}_2\left( -\frac{\eta P_t}{\sigma^2(y_m^2 + h^2)} v_1 \right) \notag \\
    &\quad- \frac{1}{\alpha} \mathrm{Li}_2\left( -\frac{\eta P_t}{\sigma^2(y_m^2 + h^2)} v_2 \right)
\Bigg].
\end{align}

\setcounter{equation}{0}
\renewcommand\theequation{C.\arabic{equation}}
\section{Proof of Lemma 3}\label{proofthree}
Without considering waveguide propagation loss, the average achievable rate can be derived using \cite[(2.733.1)]{edition2007table} as
\begin{align}
R_{p} 
&= \frac{1}{\pi r^2 \ln 2} \bigg( 
\int_{-l}^{l} \int_{-\rho(x_u)}^{\rho(x_u)} 
\ln \left(1 + \frac{\eta P_t}{\Delta_0}\right) dy_u dx_u \notag \\
&\quad+ \int_{l}^{r} \int_{-\rho(x_u)}^{\rho(x_u)} 
\ln \left(1 + \frac{\eta P_t}{\Delta_-}\right) dy_u dx_u \notag \\
&\quad+ \int_{-r}^{-l} \int_{-\rho(x_u)}^{\rho(x_u)} 
\ln \left(1 + \frac{\eta P_t}{\Delta_+}\right) dy_u dx_u
\bigg) \notag \\
&= \frac{2}{\pi r^2 \ln 2} \Bigg[\int_{0}^{l} \Bigg(
\sqrt{r^2 - x_u^2} \ln\left(r^2 - x_u^2 + h^2 + \frac{\eta P_t}{\sigma^2}\right) \notag \\
&\quad + 4\sqrt{h^2 + \frac{\eta P_t}{\sigma^2}} \arctan\left( \frac{\sqrt{r^2 - x_u^2}}{\sqrt{h^2 + \frac{\eta P_t}{\sigma^2}}} \right) \notag \\
&\quad - \sqrt{r^2 - x_u^2} \ln\left(r^2 - x_u^2 + h^2\right) \notag \\
&\quad - 4h \arctan\left( \frac{\sqrt{r^2 - x_u^2}}{h} \right)
\Bigg) dx_u \notag \\
&\quad+ \int_{l}^{r} \Bigg(
\sqrt{r^2 - y^2} \ln\left(r^2 + h^2 + l^2 - 2ly + \frac{\eta P_t}{\sigma^2} \right) \notag\\
&\quad - 2 \sqrt{r^2 - y^2} \ln\left(r^2 + h^2 + l^2 - 2ly \right) \notag \\
&\quad - 4\sqrt{h^2 + (y - l)^2} \arctan\left( \frac{\sqrt{r^2 - y^2}}{\sqrt{h^2 + (y - l)^2}} \right) \notag \\
&\quad + 4 \Lambda(y) \arctan\left( \frac{\sqrt{r^2 - y^2}}{\Lambda(y)} \right)
\Bigg) dx_u 
\Bigg].
\end{align}

Then, by applying the Gauss–Chebyshev quadrature method \cite{kumar2022performance}, a closed-form approximation of the average achievable rate can be derived as
\begin{align}
    \bar{R}_p &\approx \frac{2}{\pi r^2 \ln 2} \bigg[
\frac{l \pi}{2V} \sum_{v=1}^{V} \sqrt{1 - \zeta_v^2}  f_1(\chi_v^{(1)})\notag \\
&\quad+
\frac{(r - l) \pi}{2V} \sum_{v=1}^{V} \sqrt{1 - \zeta_v^2}  f_2(\chi_v^{(2)})
\bigg].
\end{align}

\setcounter{equation}{0}
\renewcommand\theequation{D.\arabic{equation}}
\section{Proof of Lemma 6}\label{prooffour}	
When considering partial coverage waveguide with propagation loss, the average achievable rate can be expressed as 
\begin{align}
R_{p} 
= \frac{1}{\pi r^2 \ln 2} \Bigg[
& \int_{-l}^{l} I_1(x_u)\,dx_u
+ \int_{l}^{r} I_2(x_u)\,dx_u \notag\\
& \qquad \qquad + \int_{-r}^{-l} I_3(x_u)\,dx_u
\Bigg],
\end{align}

\noindent\text{where}
\begin{align}
I_1(x_u) &\triangleq 
\int_{-\sqrt{r^2 - x_u^2}}^{\sqrt{r^2 - x_u^2}}
\ln \!\left(1 + \frac{\eta P_t e^{-\alpha(x_u + l)}}{\sigma^{2}(y_u^2 + h^2)}\right)\, dy_u, \notag\\[0.6ex]
I_2(x_u) &\triangleq 
\int_{-\sqrt{r^2 - x_u^2}}^{\sqrt{r^2 - x_u^2}}
\ln \!\left(1 + \frac{\eta P_t e^{-2 \alpha l}}{\sigma^{2}\!\big(y_u^2 + h^2 + (x_u - l)^2\big)}\right)\, dy_u, \notag\\[0.6ex]
I_3(x_u) &\triangleq 
\int_{-\sqrt{r^2 - x_u^2}}^{\sqrt{r^2 - x_u^2}}
\ln \!\left(1 + \frac{\eta P_t}{\sigma^{2}\!\big(y_u^2 + h^2 + (x_u + l)^2\big)}\right)\, dy_u. \notag
\end{align}

Then, by applying the Gauss–Chebyshev quadrature method\cite{kumar2022performance}, a closed-form approximation of the average achievable rate can be derived as
\begin{align}
    R_p \approx \frac{1}{n r^2 \ln 2} \Bigg[
    & 2l \sum_{k=1}^n F_1(x_{1, k}) + (r-l) \sum_{k=1}^n F_2(x_{2, k}) \notag \\
    & \qquad \qquad + (r-l) \sum_{k=1}^n F_3(x_{3, k})
    \Bigg].
\end{align}

\bibliographystyle{IEEEtran}
\bibliography{MyRef}

\end{document}